# Electromagnetic actuation for a vibrotactile display: Assessing stimuli complexity and usability

Michael J. Proulx [1,2], Theodoros Eracleous [1], Ben Spencer [1], Anna Passfield [1], Alexandra de Sousa [3], & Ali Mohammadi [4]

[1] Department of Psychology, University of Bath, Bath, BA2 7AY, UK
[2] REVEAL Research Centre, University of Bath, Bath, BA2 7AY, UK
[3] Centre for Health and Cognition, Bath Spa University, Bath, BA2 9BN, UK
[4] Department of Electronic and Electrical Engineering, University of Bath, Bath, BA2 7AY, UK

Corresponding Author:
Michael J. Proulx [1,2]
Department of Psychology, University of Bath, Bath, BA2 7AY, UK
Email address: m.j.proulx@bath.ac.uk

## Abstract

Sensory substitution has influenced the design of many tactile-visual substitution systems with the aim of offering visual aids for the blind. This paper focuses on whether a novel electromagnetic vibrotactile display, a four-by-four vibrotactile matrix of taxels, can serve as an aid for dynamic communication for visually impaired people. A mixed methods approach was used to firstly assess whether pattern complexity affected undergraduate participants' perceptive success; and secondly, if participants total score positively correlated with their perceived success ratings. A thematic analysis was also conducted on participants' experiences with the vibrotactile display and what methods of interaction they used. The results indicated that complex patterns were less accurately perceived than simple and linear patterns respectively, and no significant correlation was found between participants' score and perceived success ratings. Additionally, most participants interacted with the vibrotactile display in similar ways using one finger to feel one taxel at a time; arguably, the most effective strategy from previous research. This technology could have applications to navigational and communication aids for the visually impaired and road users.

# Introduction

Sensory substitution was first proposed by Bach-y-Rita (1967) in which one sensory system can perform the functions of another, potentially impaired, sensory system. This has influenced the development of many technologies that aim to facilitate the sensory perception of deaf and blind individuals through a human-machine interface (Bach-y-Rita & Kercel, 2003). A key example, and the focus of this research, being the translation of visual stimuli into tactile perceptions through an artificial stimulant. These sensory substitution systems have manifested in a variety of designs and have produced positive results in not only recognising visual forms, but also placing them in space (Lenay, Canu & Villon, 1997). Proulx, Ptito & Amedi (2014) provide a collection of reviews on the diversity of applications sensory substitution can have to improve perceptual abilities and assist those with perceptual disabilities. Overall this research would suggest that visuo-tactile sensory substitution systems can have a significant impact on developing lifestyle aids for visually impaired individuals (Proulx et al, 2016).

Research on the usefulness of sensory substitution systems predominantly focuses on assisting the visually impaired with wayfinding and spatial recognition. Apparatus using vibrotactile stimulation applied through matrices onto the back, abdomen, forehead and hand have been used; and results indicate participants can effectively avoid obstructions and navigate surroundings (Johnson & Higgins, 2006; Segond, Weiss & Sampaio, 2005; Zelek, Bromley, Asmar & Thompson, 2003). One significant limitation of these findings was identified: the technology was cumbersome and lacked portability. Furthermore, there was a specific application of navigation with little use in other fields. Therefore, the use of the vibrotactile display technology within this study, a four-by-four vibrotactile matrix that can be interacted with by hand, could offer better usability (Mohammadi, Abdelkhalek, & Sadrafshari, 2020). The use of a prototype to assess the usability of the vibrotactile display in communicating digital information as well as navigation could inspire the development of more practical sensory substitution systems with more diverse applications.

Due to the prototypical nature of the vibrotactile display, this research also aims to understand user interactions and improve the perceptual range that can be achieved by developing the design. Kaczmarek, Webster, Bach-y-Rita, and Tompkins (1991) found that areas of the skin had varying concentrations of receptors for different tactile stimuli; for example, vibrotactile receptors had higher concentration in the fingertips than in the palms. Visell (2009) concluded in their review that the design and mode of human-system interface moderated the effectiveness of substitution systems. There is some consistency that the effective tactile field of view is limited to one finger from either hand as visual acuity decreases with more points of contact (Craig, 1985; Loomis, Klatzky & Lederman, 1991). Furthermore, as Cooley (2004) claims the relationship between the individual and tactile device is essential to the perceptual experience; the participants' interactions with the vibrotactile display should be analysed to outline any preferred methods.

The primary aim is to assess whether the vibrotactile display can be used to improve the dynamic communication of image information for visually impaired individuals by representing the graphical information to their fingertips. Past work with people who are visually impaired has found that recognizing images represented by raised lines is a challenge; Kennedy (1983), for example, reported that fewer than 20% of pictures were identified by touch. However, Kennedy also found that providing a bit of contextual information, such as a single word as a caption allowed for all to be identified. This role of having some top-down understanding might play a role in those with acquired blindness performing better than the congenitally blind due to having prior conceptions about what the pictures might represent and how they are commonly represented (Heller, 1989). Although people with visual impairments are part of the lab as researchers, we chose here to focus on data collection from sighted control participants due to the low-level perceptual nature of the task as an efficient means of data collection early in the prototyping process (Brulé, Tomlinson, Metatla, Jouffrais, & Serrano, 2020). We have future co-design sessions and testing by participants with visual impairments planned for later stages when their valuable time is most beneficial for further prototyping.

Due to the challenging nature of recognising pictures by touch, even with high resolution line drawings (Kennedy, 1983), and the low resolution of the device prototype being assessed, we presented a series of relatively simple stimuli in the manner of Heller (1989). Humans have information processing capacities that are limited both by the sensory systems (Haigh et al., 2013) and attentional mechanisms (Brown & Proulx, 2016), so the patterns were selected with these limits in mind. This research will incorporate both quantitative and qualitative techniques in a mixed methods approach to determine whether participants can successfully recognise patterns and how they interacted with the device. It was hypothesised that that as the complexity of the pattern increased, the participants' perceptual success would decrease (Brown, Simpson, & Proulx, 2014). Secondly, participants' perceived success should positively correlate with their overall success on the task. Usability measures were collected to determine which aspects of the vibrotactile display could be improved in subsequent designs. Whilst evidence supports the effectiveness of using single fingertips over multiple points of contact, a qualitative analysis of participants' chosen interactive methods can be used to aid the development of the vibrotactile display design to fit the needs of the user.

## Materials & Methods
*Participants*

Nine male and seventeen female undergraduate students from Bath were recruited as participants for this study (age, M = 19.69, SD = 1.761), via opportunity sampling methods. No visually impaired participants were included on the basis that recording their responses would have required a different procedure. All participants provided informed consent and the study was approved by the University of Bath Department of Psychology Research Ethics Committee (Reference number: 0125-19-21).

*Apparatus and Materials*

The performance of vibrotactile display devices is mainly determined by the electromechanical actuators that convert the electrical signals from digital processors to vibrotactile excitations. Piezoelectric (Wang, Q. & Hayward, V., (2010)), and electromagnetic (Zárate, J. & Shea, H. (2016)), (Mohammadi, A., Abdelkhalek, M., & Sadrafshari, S. (2020)) actuators have been proven to be the most practical transducers commonly used in these display devices. Electromagnetic transducers offer larger mechanical impact, which is rather impossible with piezoelectric actuators of the same size in current technologies. Complicated transducer structures specifically designed for the commercially available vibrotactile displays increases the manufacturing costs of these products. (See the price list for Smart Beetle, Brilliant BI 14 and Focus Blue on the RNIB online shop: https://shop.rnib.org.uk/). Using off-the-shelf components and 3D printed tactile pad, significantly reduces the manufacturing costs of the vibrotactile display used in this study.

The vibrotactile display consists of a four-by-four array of 3D printed ferrmagnetic taxels on top of a matrix array of corresponding electromagnetic coils (Lui & Mohommadi, 2018). The diameter of each vibrating taxel in this prototype is 7mm, which has a circular protrusion at the middle with 4mm diameter. The centre to centre pitches of adjacent taxels are 1.25cm as shown in Figure 1. Nevertheless, the spatial resolution can be improved by the new technique presented in (Mohammadi, A., Abdelkhalek, M., & Sadrafshari, S. (2020)). The vibration frequency and amplitude are fixed at 60Hz and around 0.5mm by supplying controlled electrical signal through the electronic control unit. Though these values are adjustable by either the microcontroller code or the spring-screw system under each coil if necessary.

This prototype can be powered by 9V rechargeable batteries. In addition, the wireless connection to mobile phone handsets provided by the microcontroller (Arduino) Bluetooth module allows portable operation its the device. During the field experiments we used power adaptors to supply electrical power and interfaced the display to the lab desktop computers through USB link to focus the reliability tests only on the electromagnetic tactile interface. Further miniaturisation of the pad size to fit at the back of mobile phone handsets is of ongoing research in the Microsystems Research Lab at the University of Bath (https://www.bath.ac.uk/announcements/handheld-display-that-could-help-blind-people-perceive-information-through-touch-wins-award/).

Taxels were activated by Windows Visual Studio (2017) software on computer, though the device can be operated by android smartphones also (see Figure 1). A cardboard box was used to cover pad to prevent participants from seeing it though not restricting their access to touch it. The independent variable in this experiment was the complexity of stimulus on the vibrotactile display, consisting of three levels: simple (dots), linear (lines) and complex (patterns). The dependent variable was the number of correct identifications of the displayed shapes in each condition. A repeated measures design was used, whereby all participants were exposed to the same shapes across all three levels of complexity (5 dots, 5 lines and 7 patterns, see Appendix A) in identical order. There were five 'simple' trials containing 1-3 non-adjacent activated taxels (see Figure 2a), five 'linear' trials with 2-4 adjacent taxels forming different straight-line orientations (see Figure 2b), and seven 'complex' trials of more detailed geometric patterns, shapes or emoticons (see Figure 2c).

INSERT FIGURE 1 ABOUT HERE

INSERT FIGURE 2 ABOUT HERE

The questionnaire was designed using Qualtrics and participants recorded their responses to each trial on a representative grid (see Appendix A and B) where they select activated taxels on a separate laptop or smartphone. Reliability analyses, using Cronbach's alpha, indicated that the overall scale has strong internal reliability ($\alpha = .81$). The usability questionnaire was adapted from Hart and Staveland's (1988) NASA-TLX Task Load Index, implementing a five-point Likert scale from '1' indicating a very low score to '5' indicating a very high score. To better assess aspects of the prototype that will impact future design considerations and designs, we chose to use the NASA-TLX due to its focus on the difficulty of using the device and the load demanded of it. The NASA-TLX measures consisted of mental demand, physical demand, temporal demand, perceived success, effort, frustration, and application.

*Procedure*

Participants were instructed to feel the array of the vibrotactile display, familiarising themselves with the 4x4 layout (see Appendix A) for a couple of minutes while the researcher prepared the software interface. The vibrotactile display was placed underneath a box, wide enough for relative non-restriction but which did not allow visual access to the participant. Participants were given three practice tasks, with verbal feedback on which taxels were active, before proceeding with the trials.

Each trial began with the researcher configuring the vibrotactile display's active taxels using coordinating software on a computer connected to the vibrotactile display. Participants were instructed to touch the vibrotactile display only after the pattern was inputted, and then the researcher initiated the stimulation and the participant was allowed to identify the active taxels for a maximum of 30 seconds. They then had to remove their hand(s) from the pad, and input their answer on Qualtrics' answer box, selecting the taxels they believed were active. The researcher then changed the pattern on the vibrotactile display, beginning the next trial. A trial was marked as 'correct' if all (and only) the taxel(s) that were activated were identified.

After completing all of the identification trials, participants proceeded to answer qualitative questions in regard to the difficulty of the task in terms of physical and temporal demand, their perceived success in the task, the method they used to decipher shapes on the pad, as well as the vibrotactile display's usefulness and potential applications (see Appendix B). The vibrotactile display trials were kept close to 15 minutes due to overheating and maintenance concerns, while the whole experiment ran for an approximate total of 25 minutes. After the experiment, participants were debriefed. The research was approved by the Research Ethics Committee at the University of Bath, which follows the ethical guidelines of the British Psychological Society.

*Design*

A mixed methods approach was used including a within-groups design where the independent variable of trial complexity had three levels of simple patterns (one to three points), linear patterns (single lines), and complex patterns (multiple points arranged in shapes).

*Qualitative Data Analysis*

A thematic analysis was planned to be carried out on the open-ended question to do with participants' approaches to the task in order to identify preferred interactive techniques with the vibrotactile display. This approach was taken due to the ability to devise useful, meaningful results without prior conceptions that will enable future quantitative assessments of methods and strategies used by participants (Braun & Clarke, 2006). One author trained in the technique carried out the analysis; it is not intended to be quantitative with a comparison among multiple coders, but instead to be as source of discovering trends to direct future work (Nowell, et al., 2017). An inductive approach was used such that modes of interaction could be derived from the data which may differ from previous research on tactile perception. A realist epistemological approach would be the most appropriate in this instant as themes will be derived based on participants actual experiences with the vibrotactile display (all text data is provided in Appendix C of the Supplementary Information).

## Results

Two participants were detected as outliers in the 'Complex' condition, falling outside the 95% confidence intervals with an equal accuracy of 42.9%. However, as per Ulrich and Miller (1994), removal of data only from the upper end of a distribution produces a non-representative mean compared to the mean of the true population. Critically, studies exploring specific sample sizes (Cousineau & Chartier, 2010; Van Selst & Jolicoeur, 1994), propose decision criteria (z-scores) of 2.576 and 2.431 for a sample size of 26 respectively, which decisively indicated keeping the two participants as non-outliers.

Evaluating assumptions for Analysis of Variance (ANOVA), Mauchley's Sphericity test indicated a violation of sphericity ($\chi^2$= 7.508, p= .023). A Shapiro-Wilk test, which has been designed to most powerfully test sample sizes below 50 (Steinskog, Tjøstheim & Kvamstø, 2007), indicated violations of normality in all three complexity levels (simple: p= .001; linear: p= .015; complex: p<.001). However, Schmider et al. (2010) found that ANOVA in samples above 25 can be robust to violations of normality, hence a repeated-measures ANOVA was performed. Due to violation of sphericity, a correction was used to account for lack of sphericity whilst reducing Type I error rate. As epsilon values were $\varepsilon$>0.75 (Greenhouse-Geisser: $\varepsilon$= .788, Huynh-Feldt: $\varepsilon$=.832) the Huynh-Feldt correction was used, as the alternative Greenhouse-

Geisser test would be too conservative in rejecting the assumption of sphericity at that epsilon value, per Ginder (1992). The descriptive statistics of results are presented in Table 1.

INSERT TABLE 1 ABOUT HERE

The mean scores for each of the complexity levels were calculated and presented in Figure 3. Participants had higher mean scores for the simple block of trials compared to the linear and complex blocks, and the mean scores for the linear block were higher than that of the complex block.

INSERT FIGURE 3 ABOUT HERE

The ANOVA results indicated a significant effect of complexity on identification rates ($F(1.665, 41.613)=57.599$, $p<.001$, $\eta 2=.697$). This initiated post hoc pairwise comparisons of the three conditions, using non-parametric Wilcoxon Signed Rank test, due to the non-normality of the data, as well as a Bonferroni correction (p-value=.05/3 =.0167). The contrast results are shown in Table 2:

INSERT TABLE 2 ABOUT HERE

The mean score and standard deviation were also calculated for each of the adapted usability scales (Table 3). There were mean medium level responses for mental demand, temporal demand, perceived success, effort, and application ($2.5 < M < 3.5$); and mean low level responses were recorded for physical demand and frustration ($1.5 < M < 2.5$).

INSERT TABLE 3 ABOUT HERE

Twenty-two participants used the same method (one finger running through the pad), with the other 4 using various methods, hence the correlation between success and method of use was not examined. The correlation between actual success, calculated as the total score from all three conditions transformed score/5, with perceived success and perceived temporal demand were inspected. A Spearman's rank-order correlation indicated no significant correlation (n= 26, rs=-.096, p=.640) associating perceived success with actual success. Simultaneously, a significant relationship was indicated (n=26, rs=-.396, p=.045) associating increasing perceived temporal demand with decreasing actual success.

*Qualitative Thematic Analysis*

For the open-ended question asking about the participants' approach to the task and their mode of interface with the pad was analysed to determine any distinct techniques. One theme was identified as the most common approach to the task, which was reported by approximately

81% of the participants. While four other techniques were observed, the singular-systematic approach was clearly the most popular.

This technique comprised of using one hand and one finger to feel each individual taxel, one at a time in varying systematic orders. For those who provided specific details, this was usually done with the index finger on the participant's dominant hand, working in a top-down approach starting from the top left taxel (A1). Whilst many participants did not specify these details, the following extracts outline the emphasis on the use of a singular finger from one hand moving in a systematic progression. The following extracts are from the full report in the Supplemental Information.

*Extract 1*. (from Participant 1)
-   Feel one taxel at a time along each row

*Extract 2*. (from Participant 4)
-   Using one finger at a time (right index)

*Extract 3*. (from Participant 12)
-   Felt one taxel at a time with my right hand (dominant) – went along the lines and entered the taxel on the computer after every one

The singular-systematic approach as Extract 1 encapsulates has a certain simplicity in the assessment of tactile stimulation of a single point rather than a holistic matrix. This supports the findings of Loomis, Klatzky & Lederman, (1991) suggesting that the tactile field of view is limited to one finger at a time due to interference effects. Furthermore, Extract 2 and Extract 3 both reference their specific hand and finger used relating to handedness which may suggest that there is more sensitivity or confidence in using the dominant side.

*Extract 4*. (from Participant 24 and Participant 25)
-   One hand at a time

However, Extract 4 shows evidence of two participants who provided the same ambiguous response to the question, which makes inferring the singular-systematic approach difficult. Due to the design of the vibrotactile display, one can safely assume that this comment, and similar others, are describing an approach using one finger to feel a single taxel at a time. This is because feeling every taxel at once with the whole hand would make ascertaining the correct pattern extremely unlikely.

## Discussion

Responses to the usability questionnaire indicate that while the vibrotactile display was not physically difficult or frustrating to use, there are still aspects of the design that could be improved. Of course, this study assessed usability in the context of testing a prototype, so it is also possible there was an impact of the incidental design of the hardware on the cognitive aspects of its usability given that has an influence on the human interaction. The mental and temporal demand mean scores could indicate that some felt that it was quite challenging to interpret the patterns activated on the vibrotactile display within the given timeframe. This suggests that the vibrotactile display may require longer exposure time to ascertain the correct pattern or that the patterns themselves were quite hard to perceive and visualise. The medium level of effort required could support the latter interpretations as the complex patterns had the lowest mean score of the three blocks and this could reflect the difficulty in accurately perceiving which taxels were activated. This is supported by previous research that suggests vibrotactile displays can prove difficult for the receptors in the skin to resolve the amount of stimulation it receives, especially for complex or detailed images and patterns (Easton, 1992).

   The results of the ANOVA could support this argument as the decrease in perceptive accuracy paired with increases in pattern complexity suggests the vibrotactile display is limited to communication of simplistic information. However, a similar sized vibrotactile pad has been shown to accurately convey navigational signals and alphabetical patterns to the feet of drivers (Kim et al., 2006). Although, these linear directional signals were delivered dynamically in a stroking motion compared to the patterns which were activated on the vibrotactile display which remained constant and appeared static. The use of tracing techniques as described in Kim et al.'s (2006) study, by following the direction of activated taxels, increased participant's perceptive accuracy. Therefore, to improve the perceptive success of participants, information should be communicated through a dynamic construction of digital information (i.e., lines, patterns, letters, or symbols), such as providing animated tactile information (cf. Novich & Eagleman, 2015). Although this mode of communication could not be interpreted by an interaction with the fingertips; alternative user-system interaction sites, such as the whole hand, should be considered.

   As there was no significant positive correlation between total score and perceived success ratings, this could suggest that patterns displayed through the vibrotactile display may be quite difficult to interpret. To determine the true cause for the lack of correlation one would have to examine the participants' responses to each trial and identify the reasons for incorrect scores. This could be contributed to not perceiving one or more taxels being active, selecting an adjacent taxel by mistake, or selecting more taxels being active due to their proximity to multiple activated taxels. However, while this raises a potential limitation of the analysis it is worth noting that the vibrotactile display prototype was temperamental and the taxels did not share a uniform strength of vibration. Some taxels were particularly weak which could have contributed to incorrect scores whilst not reducing participants' perceived success. A follow up study should be conducted with a vibrotactile display that has addressed these issues to produce more valid and reliable results.

   Another point for discussion is whether the singular-systematic approach identified in the thematic analysis could have been a moderating variable of participants' score. These results could simply highlight preferred interactive methods consistent with previous findings that suggest using one finger is the limit of human tactile field of view (Loomis, Klatzky & Lederman, 1991). However, Morash, Pensky and Miele (2013) reject the idea of a tactile field of view as touch operates differently to vision and multiple fingers can be beneficial for searching tasks. This approach could be applied to the vibrotactile display when attempting to identify activated taxels, although the surface area of the pad may need to increase to accommodate for applying multiple fingers at a time. Previous applications of similar technology to navigational driving aids have produced positive results on perceptive ability when sending dynamic messages and complex symbols through the foot. Therefore, future research should focus on improving the reliability of the vibrotactile display design through uniform vibrations across taxels; and by using tracing techniques to perceive complex patterns from directional taxel activation could drastically improve perceptive accuracy. Of course, it is important to note that the current findings arise from sighted participants, and the strategies might be different for visually impaired users. This would be particularly true for those fluent in Braille-reading, though we anticipate that to be a minority among the eventual end-users given that Braille fluency is on the decline in many countries due to expense, a lack of political support and the rise of auditory accessibility, such as text-to-speech technology, even though there is evidence suggesting that Braille literacy correlates with well-being (Silverman & Bell, 2018).

While the limitations of the vibrotactile display design have been discussed with the implications on the results, the overall design of the experiment was very thorough, and the use of a mixed methods approach was very beneficial (Almalki, 2016). The mixed methods approach enabled a broad scope of data to be collected which contributes to a better understanding of its objective usability through correct responses, and participants' subjective experience and modes of interface. These data will guide the development of the vibrotactile display design and influence future research and applications of the technology. Of course it will be important to also test future designs with people in multitasking situations, such as driving, where stimulus orientation might be key for interpretation (Richardson et al., 2020), and with visually impaired persons, who sometimes indicate different biases in spatial cognition than those with visual experience (Pasqualotto & Proulx, 2012; Pasqualotto et al, 2013).

## Conclusions

To summarise, the vibrotactile display prototype may not be useful for the dynamic communication of digital information to visually impaired persons as complex patterns were less accurately perceived than simple and linear patterns respectively. Furthermore, there was no significant correlation between participants total scores and perceived success ratings potentially indicating problems in interacting with the vibrotactile display. Most participants used a single finger to feel one taxel at a time concurrent with previous research on tactile pattern detection tasks. In conclusion, the vibrotactile display technology can be developed in such a way to enable the communication of information to the visually impaired as well as drivers or cyclists.

## **Acknowledgements**

Thank you to the PS20125 team that helped with data collection for this study, and our collaborators Prof Peter Hall, Dr Simon Hayhoe, Dr Christof Lutteroth, and Prof John Ravenscroft for discussions regarding this study.

# References

Almalki, S. (2016). Integrating Quantitative and Qualitative Data in Mixed Methods Research--Challenges and Benefits. *Journal of Education and Learning*, *5*(3), 288-296.

Bach-y-Rita, P. (1967). Sensory plasticity: Applications to a vision substitution system. *Acta Neurologica Scandinavica*, *43*(4), 417-426.

Bach-y-Rita, P., & Kercel, S. W. (2003). Sensory substitution and the human–machine interface. *Trends in cognitive sciences*, *7*(12), 541-546.

Braun, V., & Clarke, V. (2006). Using thematic analysis in psychology. *Qualitative Research in Psychology, 3*(2), 77-101.

Brown, D. J., & Proulx, M. J. (2016). Audio–vision substitution for blind individuals: Addressing human information processing capacity limitations. *IEEE Journal of Selected Topics in Signal Processing, 10*(5), 924-931

Brown, D. J., Simpson, A. J. R., & Proulx, M. J. (2014). Visual objects in the auditory system via sensory substitution: How much information do we need? *Multisensory Research, 27*, 337-357. doi: 10.1163/22134808-00002462

Brulé, E., Tomlinson, B. J., Metatla, O., Jouffrais, C., & Serrano, M. (2020, April). Review of Quantitative Empirical Evaluations of Technology for People with Visual Impairments. In *Proceedings of the 2020 CHI Conference on Human Factors in Computing Systems* (pp. 1-14).

Cooley, H. R. (2004). It's all about the fit: The hand, the mobile screenic device and tactile vision. *Journal of Visual Culture, 3*(2), 133-155.

Craig, J. C. (1985). Attending to two fingers: Two hands are better than one. *Perception & Psychophysics*, *38*(6), 496-511.

Easton, R. D. (1992). Inherent problems of attempts to apply sonar and vibrotactile sensory aid technology to the perceptual needs of the blind. *Optometry and vision science: official publication of the American Academy of Optometry*, *69*(1), 3-14.

Haigh, A., Brown, D. J., Meijer, P., & Proulx, M. J. (2013). How well do you see what you hear? The acuity of visual-to-auditory sensory substitution. *Frontiers in Psychology, 4*, 330.

Hart, S. G., & Staveland, L. E. (1988). Development of NASA-TLX (Task Load Index): Results of empirical and theoretical research. *Advances in Psychology*, *52*, 139-183.

Heller, M. A. (1989). Picture and pattern perception in the sighted and the blind: the advantage of the late blind. *Perception, 18*(3), 379-389.

Johnson, L. A., & Higgins, C. M. (2006, August 30). *A navigation aid for the blind using tactile-visual sensory substitution.* Presented at the 2006 International Conference of the IEEE Engineering in Medicine and Biology Society, New York, NY, USA. Retrieved from https://ieeexplore.ieee.org/abstract/document/4463247.

Kaczmarek, K. A., Webster, J. G., Bach-y-Rita, P., & Tompkins, W. J. (1991). Electrotactile and vibrotactile displays for sensory substitution systems. *IEEE Transactions on Biomedical Engineering*, *38*(1), 1-16.

Kennedy, J. M. (1983). What can we learn about pictures from the blind? Blind people unfamiliar with pictures can draw in a universally recognizable outline style. *American Scientist, 71*(1), 19-26.

Kim, H., Seo, C., Lee, J., Ryu, J., Yu, S. B., & Lee, S. (2006, September 17). *Vibrotactile display for driving safety information*. Presented at the 2006 IEEE Intelligent Transportation Systems Conference, Toronto, Canada. Retrieved from https://ieeexplore.ieee.org/abstract/document/1706802.

Lenay, C., Canu, S., & Villon, P. (1997, August 25). *Technology and perception: The contribution of sensory substitution systems*. Presented at the Proceedings Second International Conference on Cognitive Technology Humanizing the Information Age, Aizu-Wakamatsu City, Japan. Retrieved from https://ieeexplore.ieee.org/abstract/document/617681.

Loomis, J. M., Klatzky, R. L., & Lederman, S. J. (1991). Similarity of tactual and visual picture recognition with limited field of view. *Perception, 20*(2), 167-177.

Mohammadi, A., Abdelkhalek, M., & Sadrafshari, S. (2020). Resonance frequency selective electromagnetic actuation for high-resolution vibrotactile displays. *Sensors and Actuators A: Physical, 302*, 111818.

Morash, V. S., Pensky, A. E. C., & Miele, J. A. (2013). Effects of using multiple hands and fingers on haptic performance. *Perception, 42*(7), 759-777.

Novich, S. D., & Eagleman, D. M. (2015). Using space and time to encode vibrotactile information: toward an estimate of the skin's achievable throughput. *Experimental Brain Research, 233*(10), 2777-2788.

Nowell, L. S., Norris, J. M., White, D. E., & Moules, N. J. (2017). Thematic analysis: Striving to meet the trustworthiness criteria. *International Journal of Qualitative Methods, 16*(1), 1609406917733847.

Pasqualotto, A., & Proulx, M. J. (2012). The role of visual experience for the neural basis of spatial cognition. *Neuroscience & Biobehavioral Reviews, 36*, 1179-1187.

Pasqualotto, A., Spiller, M.J., Jansari, A., & Proulx, M. J. (2013). Visual experience facilitates allocentric spatial representation. *Behavioural Brain Research, 236,* 175-179.

Proulx, M. J., Gwinnutt, J., Dell'Erba, S., Levy-Tzedek, S., de Sousa, A. A., & Brown, D. J. (2016). Other ways of seeing: from behavior to neural mechanisms in the online "visual" control of action with sensory substitution. *Restorative Neurology & Neuroscience, 34*, 29-44.

Proulx, M. J., Ptito, M., & Amedi, A. (2014). Multisensory integration, sensory substitution and visual rehabilitation. *Neuroscience and Biobehavioural Reviews, 41,* 1-2.

Richardson, M., Esenkaya, T., Petrini, K., & Proulx, M. J. (2020). Reading with the tongue: Perceiving ambiguous stimuli with the BrainPort. In *Proceedings of the 2020 CHI Conference on Human Factors in Computing Systems.* (pp. in press). ACM.

Segond, H., Weiss, D., & Sampaio, E. (2005). Human spatial navigation via a visuo-tactile sensory substitution system. *Perception, 34*(10), 1231-1249.

Silverman, A. M., & Bell, E. C. (2018). The association between Braille reading history and well-being for blind adults. *Journal of Blindness Innovation & Research, 8*(1), https://www.nfb.org/images/nfb/publications/jbir/jbir18/jbir080103.html.

Visell, Y. (2009). Tactile sensory substitution: Models for enaction in HCI. *Interacting with Computers*, *21*(1-2), 38-53.

Wang, Q. & Hayward, V. (2010). Biomechanically optimized distributed tactile transducer based on lateral skin deformation. *Int. J. Robotics Res. 29* (4), 323–335.

Zárate, J. & Shea, H. (2016). Using pot-magnets to enable stable and scalable electromagnetic tactile displays. *IEEE Trans. Haptics 10* (1), 106–112.

Zelek, J. S., Bromley, S., Asmar, D., & Thompson, D. (2003). A haptic glove as a tactile-vision sensory substitution for wayfinding. *Journal of Visual Impairment and Blindness*, *97*(10), 621-632.

# Figure 1

The graphical user interface controlled by the experimenter on a PC screen and the vibrotactile apparatus. The vibrotactile display prototype, using electromagnetic actuation to present stimuli of varying complexity in the experimental trials.

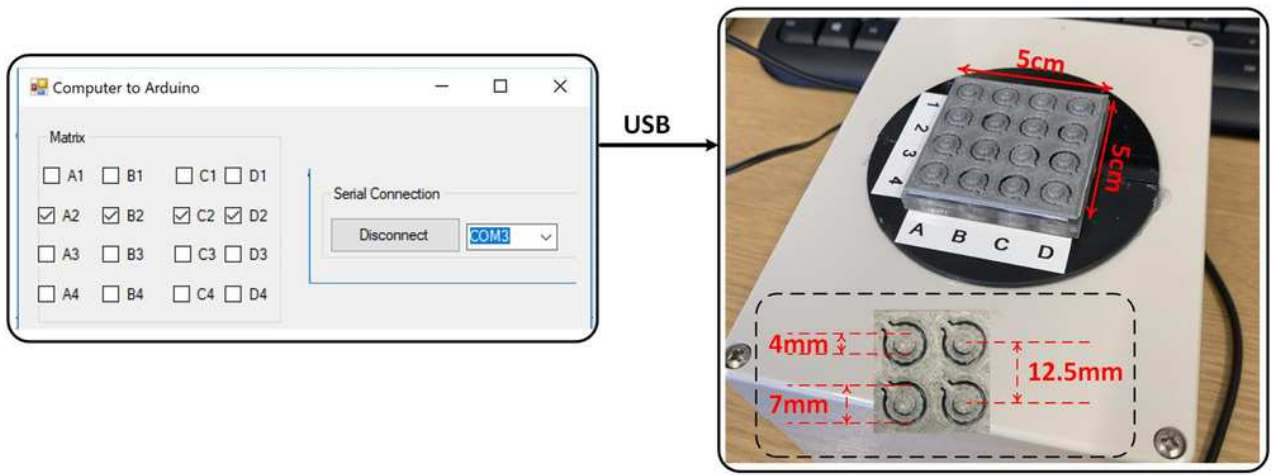

# Figure 2

Examples of the stimuli configuration types (all trial types are available in the Supplemental Information).

(A) Example of simple trial. (B) Example of a linear trial. (C) Example of a complex trial.

| A | A | B | C | D |
|---|---|---|---|---|
| 1 |   |   |   |   |
| 2 |   |   |   |   |
| 3 | x |   |   |   |
| 4 |   |   | x |   |

| B | A | B | C | D |
|---|---|---|---|---|
| 1 | x |   |   |   |
| 2 | x |   |   |   |
| 3 | x |   |   |   |
| 4 | x |   |   |   |

| C | A | B | C | D |
|---|---|---|---|---|
| 1 |   |   |   |   |
| 2 |   |   |   | x |
| 3 | x |   | x |   |
| 4 |   | x |   |   |

# Figure 3

Violin plot showing the distributions of the mean scores for each block of trials. The Simple condition performance was significantly different from the Linear and Complex conditions, p < .001.

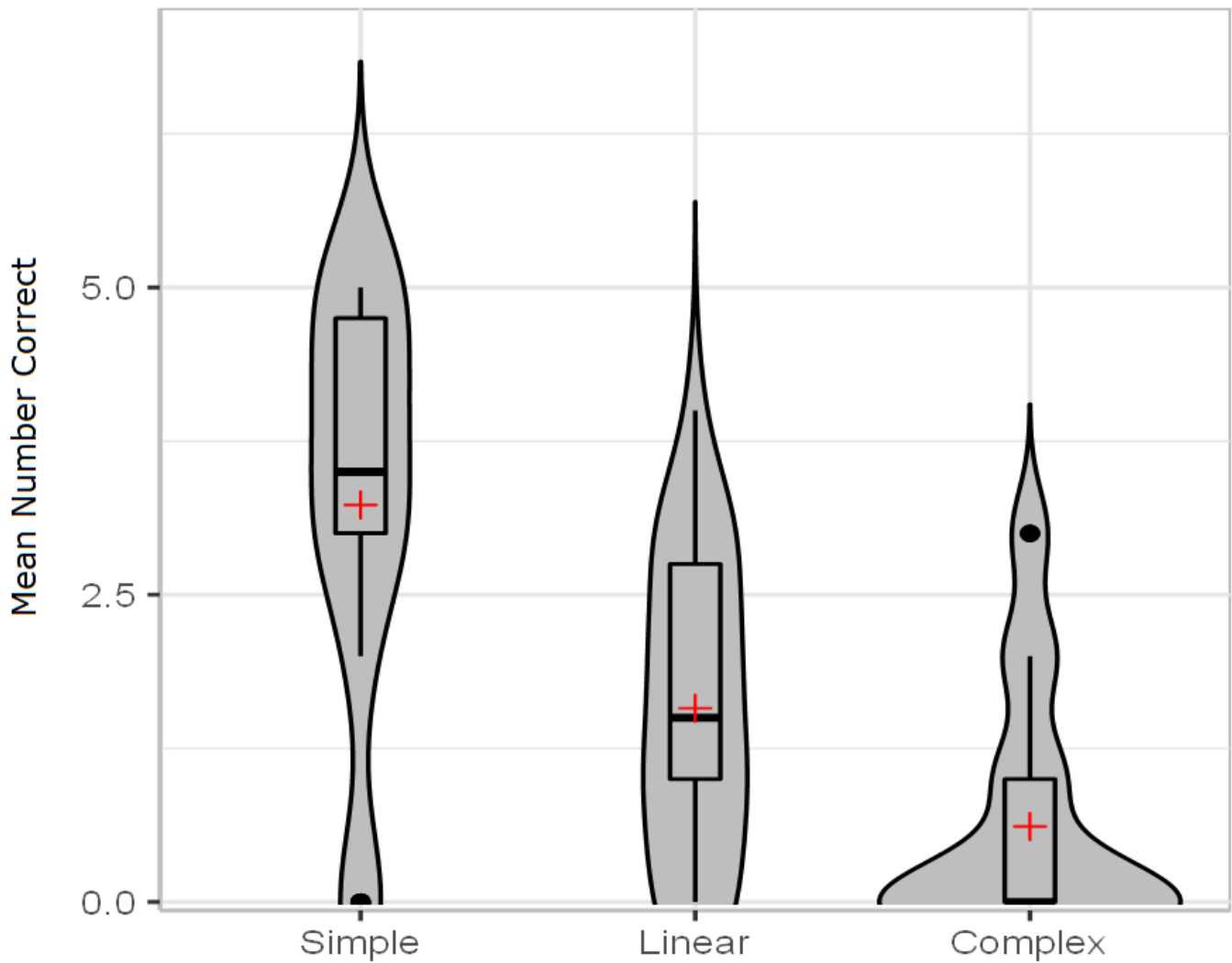

# Table 1(on next page)

*Descriptive statistics of identification percentage scores in three levels of Complexity.*

Table 1. *Descriptive statistics of identification percentage scores in three levels of Complexity*

| Complexity | N | Mean | Std. Deviation | Mode | Mode Frequency |
|---|---|---|---|---|---|
| Simple | 26 | 64.62 | 33.62 | 100 | 7 |
| Linear | 26 | 31.52 | 24.11 | 20 | 7 |
| Complex | 26 | 8.79 | 14.04 | 0 | 17 |

# Table 2(on next page)

*Wilcoxon Signed Rank tests across complexity levels*

Table 2. *Wilcoxon Signed Rank tests across complexity levels*

| Complexity-pairs | Z-score | p-value |
|---|---|---|
| Simple-Linear | -3.790 | <.001 |
| Linear-Complex | -3.927 | <.001 |
| Simple-Complex | -4.183 | <.001 |

# Table 3

*Mean scores and standard deviations for usability measures.*

Table 3. *Mean scores and standard deviations for usability measures.*

| Usability measure | Mean Score | Standard deviation |
|---|---|---|
| Mental demand | 2.65 | .85 |
| Physical demand | 1.54 | .95 |
| Temporal demand | 2.50 | 1.14 |
| Perceived success | 3.27 | 1.19 |
| Effort | 3.04 | 1.04 |
| Frustration | 2.27 | 1.54 |
| Application | 2.54 | .99 |